\documentclass[aps,graphicx,twocolumn,prd,eqsecnum,showpacs]{revtex4-1}
\usepackage{graphicx}% Include figure files
\usepackage{dcolumn}% Align table columns on decimal point
\usepackage{bm}% bold math
\begin{document}

\title{Pion form factor and reactions $e^+e^-\to\omega\pi^0$ and
$e^+e^-\to\pi^+\pi^-\pi^+\pi^-$ at energies up to 2-3 GeV in the
many-channel approach.}
\author{N.~N.~Achasov}
\email[]{achasov@math.nsc.ru} \affiliation{Laboratory of
Theoretical Physics, S.~L.~Sobolev Institute for Mathematics,
630090, Novosibirsk, Russian Federation
}%
\author{A.~A.~Kozhevnikov}
\email[]{kozhev@math.nsc.ru} \affiliation{Laboratory of
Theoretical Physics, S.~L.~Sobolev Institute for Mathematics, and
Novosibirsk State University, 630090, Novosibirsk, Russian
Federation}

\date{\today}
\begin{abstract}
Using the  field-theory-inspired  expression for the pion
electromagnetic form factor $F_\pi$, a good description of the
data in the range $-10<s<1$ GeV$^2$ is obtained upon taking into
account the pseudoscalar-pseudoscalar (PP) loops. When the
vector-pseudoscalar (VP) and the axial vector-pseudoscalar (AP)
loops  are taken into account in addition to the PP ones, a good
description of the $BABAR$ data on the reaction
$e^+e^-\to\pi^+\pi^-$ is obtained at energies up to 3 GeV. The
inclusion of the VP and AP loops demands the treatment of the
reactions $e^+e^-\to\omega\pi^0$ and
$e^+e^-\to\pi^+\pi^-\pi^+\pi^-$. This task is performed with the
SND data on $\omega\pi^0$ production and the $BABAR$ data on
$\pi^+\pi^-\pi^+\pi^-$ production, both in $e^+e^-$ annihilation,
by taking into account $\rho(770)$ and the heavier $\rho(1450)$,
$\rho(1700)$, and $\rho(2100)$ resonances. The problems arising
from including of the VP and AP loops are pointed out and
discussed.
\end{abstract}

\maketitle

\section{Introduction}
\label{intro}~

Some time ago we suggested a new expression for the
electromagnetic form factor of the pion $F_\pi$
\cite{ach11,ach12,ach13}, which describes the data on the reaction
$e^+e^-\to\pi^+\pi^-$ \cite{snd,cmd,kloe,babar} restricted to the
time-like region $4m^2_\pi<s\leq1$ GeV$^2$. The expression takes
into account the strong resonance  mixing via common decay modes
and the $\rho\omega$ mixing. It has both the correct analytical
properties and the normalization condition $F_\pi(0)=1$, and can
be represented in the form:
\begin{widetext}
\begin{eqnarray}
F_\pi(s)&=&\frac{1}{\Delta}(g_{\gamma\rho_1},g_{\gamma\rho_2},g_{\gamma\rho_3},...)\left(%
\begin{array}{ccccc}
  g_{11} & g_{12} & g_{13}&... & \frac{g_{11}\Pi_{\rho_1\omega}}{D_\omega} \\
  g_{12} & g_{22} & g_{23} &...& \frac{g_{12}\Pi_{\rho_1\omega}}{D_\omega} \\
  g_{13} & g_{23} & g_{33} &...& \frac{g_{13}\Pi_{\rho_1\omega}}{D_\omega} \\
   ...   & ...    & ...    & ...&...\\
  \frac{g_{11}\Pi_{\rho_1\omega}}{D_\omega} & \frac{g_{12}\Pi_{\rho_1\omega}}{D_\omega} &
  \frac{g_{13}\Pi_{\rho_1\omega}}{D_\omega} &...& \frac{\Delta}{D_\omega} \\
\end{array}%
\right)\left(%
\begin{array}{c}
  g_{\rho_1\pi\pi} \\
  g_{\rho_2\pi\pi} \\
  g_{\rho_3\pi\pi} \\
  ... \\
  0 \\
\end{array}%
\right),\label{Fpi}\end{eqnarray}\end{widetext}where $i$
($i=1,2,3,...$) counts the $\rho$-like resonance  states
$\rho_1\equiv\rho(770)$, $\rho_2\equiv\rho(1450)$,
$\rho_3\equiv\rho(1700)$, ..., the quantity
\begin{equation}
g_{\gamma V}=\frac{m^2_V}{g_V},\label{garhg}\end{equation}
($V=\rho_{1,2,3,...},\omega$) is introduced in such a way that
$eg_{\gamma V}$ is the $\gamma V$ transition amplitude, where $e$
is the electric charge. As usual, the coupling constant $g_V$ is
calculated from the electronic width
\begin{equation}
\Gamma_{V\to
e^+e^-}=\frac{4\pi\alpha^2m_V}{3g^2_V}\label{gamee}\end{equation}
of the resonance $V$. The quantities $g_{ij}/\Delta$ are the
matrix elements of the matrix $G^{-1}$ given by Eq.~(\ref{G})
below, and $\Delta={\rm det}G$. Ellipses mean additional states
like $\rho(2100)$ etc. It is assumed that the direct
G-parity-violating decay $\omega\to\pi^+\pi^-$ is absent, that is,
$g_{\omega\pi\pi}=0$. The quantity $\Pi_{\rho_1\omega}$ is
responsible for the $\rho\omega$ mixing. See Ref.~\cite{ach11} for
more details concerning Eq.~(\ref{Fpi}). We note  that an
expression similar to Eq.~(\ref{Fpi}) was used earlier
\cite{ach97} for the description of  data in the time-like domain,
but it had a disadvantage in that the normalization condition
$F_\pi(0)=1$ was satisfied only within an accuracy of 20$\%$.

Using the resonance parameters found from fitting the data
\cite{snd,cmd,kloe,babar}, the continuation to the space-like
region $s<0$ was made, and the curve describing the  behavior of
$F_\pi(s)$ in the range $-0.2\mbox{ GeV}^2<s<0$ GeV$^2$ was
obtained \cite{ach11} and compared with the data \cite{amendolia}
in this interval of the momentum transfer squared. The space-like
interval was further expanded to $s=-10$ GeV$^2$ in  a subsequent
work \cite{ach13},  and a comparison was made with the data
\cite{bebek,horn,tadev} existing in that interval. The basic
ingredient in the above treatment is the inclusion of the
pseudoscalar-pseudoscalar  (PP) loops, specifically, the
$\pi^+\pi^-$ and $K\bar K$ ones. These contributions are dominant
at the center-of-mass energy $\sqrt{}s\leq1$ GeV. Going to higher
energies (up to 3 GeV) of the reaction $e^+e^-\to\pi^+\pi^-$
\cite{babar} - requires the inclusion of the vector-pseudoscalar
(VP) and  axial vector-pseudoscalar (AP) intermediate states. This
is the aim of the present work. The particular VP state
$\omega\pi^0$ produced in $e^+e^-$ annihilation was studied by the
SND team in Ref.~\cite{snd13}, while the AP state of the type
$a_1\pi$ is the intermediate state in the reaction
$e^+e^-\to\pi^+\pi^-\pi^+\pi^-$ studied by $BABAR$ \cite{bab4pi}.
An attempt to describe these reactions in the framework of the
three-channel approach, taking into account the PP, VP, and AP
intermediate states, is also undertaken in the present work.
Furthermore, a suitable scheme with three subtractions for the
nondiagonal polarization operators is used in the present work as
opposed to Refs.~\cite{ach11,ach12,ach13} where the scheme with
two subtractions was used.

Of course, there are many works in the current literature devoted
to analyzing of the pion form factor in models that are different
from that proposed here and in Refs.~\cite{ach11,ach12,ach13}. In
particular,  a model with a broken hidden local symmetry added to
the $\pi^+\pi^-$ and $K\bar K$ loops at energies $\sqrt{s}\leq1$
GeV was used in Refs.\cite{ben13,ben12,ben01}, without attempting
to extend the analysis  to higher energies. A subtraction scheme
different from ours was used there for the calculation of the
pseudoscalar-loop contribution. The task of extending the energy
region above 1 GeV was undertaken in Ref.~\cite{czyz10}, taking
into account the contributions of heavier rho-like resonances.
However, the mixing among these resonances -- necessarily arising
due to their common decay modes -- was not taken into account in
that work. A model similar to the $K$-matrix approach but with
improved analytical properties was proposed in Ref.~\cite{han}. As
opposed to the above works, the present work uses the
field-theory-inspired approach to the problem which takes into
account relevant PP-, VP-, and AP-loop contributions and the
strong mixing of the $\rho$-like resonances arising via their
common decay modes.

The  paper is organized as follows. The polarization operators
arising   due to the vector - pseudoscalar and axial vector -
pseudoscalar loops (the diagonal and nondiagonal) and the
nondiagonal pseudoscalar - pseudoscalar polarization operator, are
calculated in Sec.~\ref{VPAP}. The expression for the
pseudoscalar-pseudoscalar diagonal polarization operator
\cite{ach11,ach12,ach13} is reviewed  in the same section. The
quantities for comparison with experimental data are discussed in
Sec.~\ref{quantit}. The results of the data fitting are
represented in Sec.~\ref{results}. This section also contains a
discussion of the problems that arise when including the VP and AP
loops. Section \ref{concl} contains the conclusions drawn from the
present study.

\section{Polarization operators due to pseudoscalar-pseudoscalar,
vector-pseudoscalar, and axial vector-pseudoscalar loops}
\label{VPAP}~

The final states $\pi^+\pi^-$, $\omega\pi^0$, and
$\pi^+\pi^-\pi^+\pi^-$ considered in the present work, have the
isotopic spin $I=1$. Hence, they are produced in $e^+e^-$
annihilation via the unit spin $\rho$-like  intermediate states
$\rho_1\equiv\rho(770)$, $\rho_2\equiv\rho(1450)$,
$\rho_3\equiv\rho(1700)$, etc. These states have rather large
widths and are mixed via their common decay modes. The finite
width and  mixing effects are taken into account by means of the
diagonal and nondiagonal polarization operators
$\Pi_{\rho_i\rho_j}$ \cite{ach11}. In particular, the effects of
finite width appear in the inverse propagator of the resonance
$\rho_i$ via the replacement
\begin{equation}
m^2_{\rho_i}-s\to m^2_{\rho_i}-s-\Pi_{\rho_i\rho_i}\equiv
D_i.\label{repl}
\end{equation} Indeed,
according to unitarity relation, the particular contribution to
the imaginary part of the diagonal polarization operator is due to
the real intermediate state $ab$:
\begin{eqnarray}
\frac{\Pi^{(ab)}_{\rho_i\rho_i}(s)}{s}&=&\frac{1}{\pi}
\int_{(m_a+m_b)^2}^\infty\frac{\sqrt{s^\prime}\Gamma_{\rho_i\to
ab}(s^\prime)}{s^\prime(s^\prime-s-i\varepsilon)}ds^\prime,\nonumber\\
{\rm Im}\Pi^{(ab)}_{\rho_i\rho_i}(s)&=&\sqrt{s}\Gamma_{\rho_i\to
ab}(s).
\end{eqnarray}
Hereafter, the quantity $s$ is the energy squared. As  explained
earlier \cite{ach11}, the dispersion relation written for the
polarization operator divided by $s$, automatically guarantees the
correct normalization of the form factor $F_\pi(0)=1$. In the
present work, the states which are taken into account are the PP
states $\pi^+\pi^-$, $K^+K^-+K^0\bar K^0$ of the pair of
pseudoscalar mesons, the  VP states $\omega\pi^0$,
$K^{\ast+}K^-+K^{\ast0}\bar K^0+K^{\ast-}K^++\bar K^{\ast 0}K^0$,
and the  AP states $a_1^+(1260)\pi^-+a_1^-(1260)\pi^+$,
$K_1(1270)\bar K+$ c.c.  The polarization operators due to the PP
loops are considered in detail elsewhere \cite{ach11}.

The following subtraction scheme is used in the present work. The
diagonal polarization operators $\Pi_{\rho_i\rho_i}(s)$ are
regularized by making two subtractions,   at $s=0$ and at the
respective mass squared $s=m^2_{\rho_i}$, $i=1,2,...$:
\begin{equation}\label{regdiag}
{\rm Re}\Pi_{\rho_i\rho_i}(0)={\rm
Re}\Pi_{\rho_i\rho_i}(m^2_{\rho_i})=0.\end{equation} The
nondiagonal polarization operators $\Pi_{\rho_i\rho_j}(s)$,
$i\not=j$, are regularized by  making three subtractions,   at
$s=0$ at $s=m^2_{\rho_i}$, and at $s=m^2_{\rho_j}$ $i,j=1,2,...$:
\begin{equation}\label{regnondiag}
{\rm Re}\Pi_{\rho_i\rho_j}(0)={\rm
Re}\Pi_{\rho_i\rho_j}(m^2_{\rho_i})={\rm
Re}\Pi_{\rho_i\rho_j}(m^2_{\rho_j})=0.\end{equation} The
corresponding expression, in the case of two-particle state $ab$,
is
\begin{equation}\label{Piij}
\Pi^{(ab)}_{\rho_i\rho_j}(s)=g_{\rho_iab}g_{\rho_jab}\Pi_{\rho_i\rho_j}(s,m_{\rho_i},m_{\rho_j},m_a,m_b),
\end{equation}where
\begin{widetext}
\begin{eqnarray}\label{Piij1}
\Pi_{\rho_i\rho_j}(s,m_{\rho_i},m_{\rho_j},m_a,m_b)&=&\frac{s}{m^2_{\rho_i}-m^2_{\rho_j}}
\left[\frac{{\rm
Re}G^{(ab)}(m^2_{\rho_i})}{m^2_{\rho_i}}(m^2_{\rho_j}-s)-\frac{{\rm
Re}G^{(ab)}(m^2_{\rho_j})}{m^2_{\rho_j}}(m^2_{\rho_i}-s)\right]+G^{(ab)}(s),\end{eqnarray}
\end{widetext}
while $G^{(ab)}(s)\equiv G^{(ab)}(s,m_a,m_b)$,
\begin{equation}\label{Gab}
G^{(ab)}(s,m_a,m_b)=\frac{s^3}{\pi
g^2_{\rho_iab}}\int_{(m_a+m_b)^2}^\infty\frac{\sqrt{s^\prime}\Gamma_{\rho_iab}(s^\prime)ds^\prime}{s^{\prime3}(s^\prime-s-i0)},\end{equation}
with $g_{\rho_iab}$ and $\Gamma_{\rho_iab}$ being the coupling
constant and the partial width of the decay $\rho_i\to
ab$,respectively. The specific expressions for $G^{(ab)}$ and
other necessary quantities are given below. Note that a different
scheme with two subtractions for the nondiagonal PP polarization
operators was used in Ref.~\cite{ach11,ach12,ach13}.

\subsection{Pseudoscalar-pseudoscalar loop}
~

The diagonal polarization operators due to the PP loop are
represented in the form
\begin{equation}\label{PiPP}
\Pi_{\rho_i\rho_i}^{(PP)}=g^2_{\rho_iPP}\Pi^{(PP)}.\end{equation}
The function $\Pi^{(PP)}$ is
\begin{equation}\label{PiPP1}
\Pi^{(PP)}\equiv\Pi^{(PP)}(s,m_V,m_P)=\Pi^{(PP)}_0+\Pi^{(PP)}_1,\end{equation}
where
\begin{eqnarray}
\Pi^{(PP)}_0&=&\frac{s}{48\pi^2}\left[8m^2_P\left(\frac{1}{m^2_V}-\frac{1}{s}\right)
+v^3_P(m^2_V)\times\right.\nonumber\\&&\left.
\ln\frac{1+v_P(m^2_V)}{1+v_P(m^2_V)}\theta(m_V-2m_P)
-\right.\nonumber\\&&\left.
2\bar v^3_P(m^2_V)\arctan\frac{1}{\bar v_P}\theta(2m_P-m_V)\right],\nonumber\\
\Pi^{(PP)}_1&=&\frac{s}{48\pi^2}\left\{\theta(s-4m^2_P)v^3_P(s)\times\right.\nonumber\\&&
\left.\left[i\pi-\ln\frac{1+v_P(s)}{1-v_P(s)}\right]
+\right.\nonumber\\&&\left.2\theta(4m^2_P-s)\theta(s)\bar
v^3_P(s)\arctan\frac{1}{\bar v_P(s)}
-\right.\nonumber\\&&\left.\theta(-s)v^3_P(s)\ln\frac{v_P(s)+1}{v_P(s)-1}\right\},
\end{eqnarray}
\cite{fn2} and $$v_P(s)=\sqrt{1-\frac{4m^2_P}{s}},$$ $$\bar
v_P(s)=\sqrt{\frac{4m^2_P}{s}-1},$$ $\theta$ is the step function.

The function $G^{(PP)}(s)$ [Eq.~(\ref{Gab})]  necessary for the
evaluation of the nondiagonal polarization operator due to the PP
loop is
\begin{widetext}
\begin{eqnarray}\label{GPP}
G^{(PP)}(s)&=&\frac{1}{24\pi^2}\left\{\frac{s}{4m^2_P}\left[\frac{2}{15}+\frac{2}{3}v_P^2(s)-v^4_P(s)\right]
-\frac{1}{2}\theta(-s)v^3_P(s)\ln\frac{v_P(s)+1}{v_P(s)-1}+
\theta(4m^2_P-s)\theta(s)\bar{v}_P(s)\times\right.\nonumber\\&&\left.\arctan\frac{1}{\bar
v_P(s)}+\frac{1}{2}\theta(s-4m^2_P)v^3_P(s)\left[i\pi-\ln\frac{1+v_P(s)}{1-v_P(s)}\right]\right\}.
\end{eqnarray}
\end{widetext}

\subsection{Vector-pseudoscalar  loop}
~

The diagonal polarization operators due to the VP loop are
represented in the form
\begin{equation}
\Pi_{\rho_i\rho_i}^{(VP)}=g^2_{\rho_iVP}\Pi^{(VP)},\end{equation}
where the function
$\Pi^{(VP)}\equiv\Pi^{(VP)}(s,m_{\rho_i},m_V,m_P)$ is calculated
from the dispersion relation
\begin{eqnarray}
\frac{\Pi^{(VP)}}{s}&=&\frac{1}{12\pi^2}
\int_{(m_V+m_P)^2}^\infty\frac{q^3_{VP}(s^\prime,m_V,m_P)}{\sqrt{s^\prime}(s^\prime-s-i\varepsilon)}\times
\nonumber\\&&\left(\frac{s_0+m^2_{\rho_i}}{s_0+s^\prime}\right)ds^\prime.
\label{dispVP}
\end{eqnarray}The notations are as follows.  The quantity
\begin{eqnarray}\label{qab}
q_{ab}(s,m_a,m_b)&=&\left[s^2-2(m_a^2+m_b^2)+\right.\nonumber\\&&\left.
(m_a^2-m_b^2)^2\right]^{1/2}/2\sqrt{s}\label{q}\end{eqnarray} is
the momentum of the particle $a$ or $b$ in the rest frame of the
decaying particle with the invariant mass $\sqrt{s}$;
$m_{\rho_i}$, $m_V$ and $m_P$ are, respectively, the masses of the
resonance $\rho_i$, and the vector $V$ and pseudoscalar $P$ mesons
propagating in the loop, $g_{\rho_iVP}$ is the coupling constant
of the resonance $\rho_i$ with the VP state. It is well known that
the partial width of the decay $\rho_i\to VP$,
$$\Gamma_{\rho_iVP}(s)=\frac{g_{\rho_iVP}^2}{12\pi}q^3_{VP}(s,m_V,m_P),$$
grows as the energy increases. This growth spoils the convergence
of the integral (\ref{dispVP}). This is the reason for the
appearance of the function
$(s_0+m^2_{\rho_i})/(s^\prime+m^2_{\rho_i})$, in the integrand of
Eq.~(\ref{dispVP}). It suppresses the fast growth of the partial
width and improves the convergence of the above integral at large
$s^\prime$. However, the integral still remains logarithmically
divergent, and one should perform the subtraction of the real part
Re$\Pi_{\rho_i\rho_i}^{(VP)}/s$ at $s=m^2_{\rho_i}$.

\begin{figure}
  \includegraphics[width=7cm]{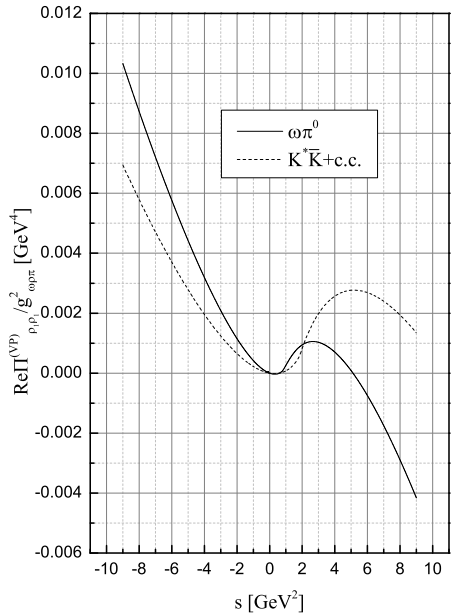}
  \caption{\label{repivp}The dependence of Re$\Pi^{(VP)}_{\rho_1\rho_1}(s)/g^2_{\omega\rho_1\pi}$ on energy squared;
  $s_0=0.09$ GeV$^2$.
  }
\end{figure}

The expression for $\Pi^{(VP)}$ resulting from Eq.~(\ref{dispVP})
can be represented in the form
\begin{equation}\label{PiVP}
\Pi^{(VP)}=\frac{1}{96\pi^2}\left[\Pi^{(VP)}_0+\Pi^{(VP)}_1+\Pi^{(VP)}_2\right],
\end{equation}
where
\begin{widetext}
\begin{eqnarray}
\Pi^{(VP)}_0&=&\frac{m^2_{\rho_i}+s_0}{s_0^2}\left(1-\frac{s}{m^2_{\rho_i}}\right)
\left\{(m_+m_-)^3\left[1-\frac{s_0}{s}\left(1+\frac{s}{m^2_{\rho_i}}\right)\right]\ln\frac{m_V}{m_P}+
m_+m_-\times\right.\nonumber\\&&\left.
\left[\frac{3}{2}(m^2_++m^2_-)\ln\frac{m_V}{m_P}+m_+m_-\right]s_0
-\frac{sm^2_{\rho_i}}{(s+s_0)(m^2_{\rho_i}+s_0)}
\left[(m^2_++s_0)(m^2_-+s_0)\right]^{3/2}\times\right.\nonumber\\&&\left.\ln\frac{\sqrt{m^2_++s_0}+\sqrt{m^2_-+s_0}}
{\sqrt{m^2_++s_0}-\sqrt{m^2_-+s_0}}\right\},\nonumber\\
\Pi^{(VP)}_1&=&-\frac{s}{m^4_{\rho_i}}\left|(m^2_+-m^2_{\rho_i})(m^2_{\rho_i}-m^2_-)\right|^{3/2}\left[
2\theta(m_+-m_{\rho_i})\theta(m_{\rho_i}-m_-)\arctan\sqrt{\frac{m^2_{\rho_i}-m^2_-}{m^2_+-m^2_{\rho_i}}}-\right.\nonumber\\&&\left.
\theta(m_{\rho_i}-m_+)\ln\frac{\sqrt{m^2_{\rho_i}-m^2_-}+\sqrt{m^2_{\rho_i}-m^2_+}}
       {\sqrt{m^2_{\rho_i}-m^2_-}-\sqrt{m^2_{\rho_i}-m^2_+}}+\theta(m_--m_{\rho_i})\ln\frac{\sqrt{m^2_+-m^2_{\rho_i}}+\sqrt{m^2_--m^2_{\rho_i}}}
       {\sqrt{m^2_+-m^2_{\rho_i}}-\sqrt{m^2_--m^2_{\rho_i}}}\right],\nonumber\\
\Pi^{(VP)}_2&=&\frac{m^2_{\rho_i}+s_0}{s(s+s_0)}\left|(m^2_+-s)(m^2_--s)\right|^{3/2}\left[\theta(m^2_--s)
\ln\frac{\sqrt{m^2_+-s}+\sqrt{m^2_--s}}{\sqrt{m^2_+-s}-\sqrt{m^2_--s}}
+\right.\nonumber\\&&\left.2\theta(s-m^2_-)\theta(m^2_+-s)\arctan\sqrt{\frac{s-m^2_-}{m^2_+-s}}+\theta(s-m^2_+)
\left(i\pi-\ln\frac{\sqrt{s-m^2_-}+\sqrt{s-m^2_+}}{\sqrt{s-m^2_-}-\sqrt{s-m^2_+}}\right)
\right],
\end{eqnarray}
\end{widetext}
while $m_\pm=m_V\pm m_P$, and $\theta$ is the usual step function.
The dependence of Re$\Pi^{(VP)}_{\rho_1\rho_1}(s)$ on energy
squared at $s_0=0.09$ GeV$^2$ is shown in Fig.~\ref{repivp}.

\begin{figure}
  \includegraphics[width=7cm]{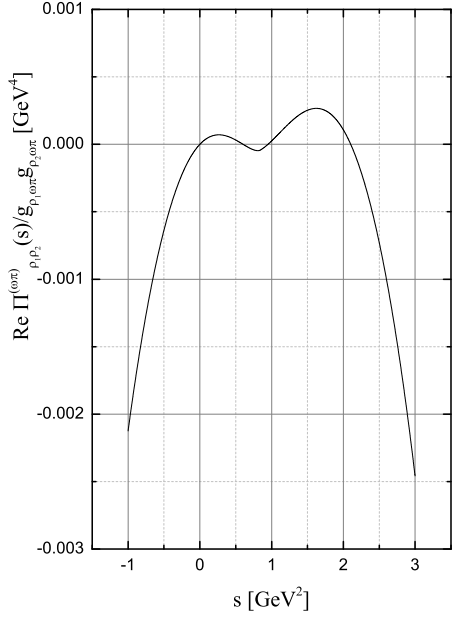}
  \caption{\label{repivpij}The dependence of Re$\Pi^{(\omega\pi)}_{\rho_1\rho_2}(s)/g_{\omega\rho_1\pi}g_{\omega\rho_2\pi}$ on energy squared in case of
  $m_{\rho_1}=770$ MeV and $m_{\rho_2}=1450$ MeV.}
\end{figure}

The function $G^{(VP)}(s)$ [Eq.~(\ref{Gab})]  necessary for the
evaluation of the nondiagonal polarization operator due to the VP
loop is
\begin{equation}\label{GVP}
G^{(VP)}(s)=\frac{s^2}{48\pi^2}\left(\frac{4m_Vm_P}{m_+m_-}\right)^3I^{(VP)}(s),
\end{equation}
where
\begin{widetext}
\begin{eqnarray}\label{IVP}
I^{(VP)}(s)&=&\left\{\frac{s^3}{16}+\frac{m^2_+}{4m_Vm_P}(s-m^2_-)\left[\frac{3s^2}{8}-\frac{m^2_+s(s-m^2_-)}{8m_Vm_P}
-\left(\frac{m_+m_-}{4m_Vm_P}\right)^2(m^2_+-s)(s-m^2_-)\right]\right\}\times\nonumber\\&&
\frac{1}{2s^3}\ln\frac{m_V}{m_P}+\frac{m_+m_-}{4m_Vm_Ps^3}\left\{\frac{s^2m^2_-}{4m_Vm_P}\left[\frac{s}{24}+
\frac{m^2_+(s-m^2_-)}{16m_Vm_P}+\frac{sm^2_-}{24m_Vm_P}\right]-\frac{s^3}{16}-\frac{sm^2_+(s-m^2_-)}{4m_Vm_P}\right.\times
\nonumber\\&&\left.
\left[\frac{3s}{8}-\frac{m^2_+(s-m^2_-)}{8m_Vm_P}\right]\right\}+\left(\frac{m_+m_-}{4m_Vm_Ps}\right)^3\left|(s-m^2_+)(s-m^2_-\right|^{3/2}
\times\nonumber\\&&\left[\frac{1}{2}\theta(m^2_--s)\ln\frac{\sqrt{m^2_+-s}+\sqrt{m^2_--s}}{\sqrt{m^2_+-s}-\sqrt{m^2_--s}}
+\theta(s-m^2_-)\theta(m^2_+-s)\arctan\sqrt{\frac{s-m^2_-}{m^2_+-s}}+\right.\nonumber\\&&\left.\frac{1}{2}\theta(s-m^2_+)
\left(i\pi-\ln\frac{\sqrt{s-m^2_-}+\sqrt{s-m^2_+}}{\sqrt{s-m^2_-}-\sqrt{s-m^2_+}}\right)\right].
\end{eqnarray}
\end{widetext}
The dependence of
Re$\Pi^{(\omega\pi)}_{\rho_1\rho_2}/g_{\omega\rho_1\pi}g_{\omega\rho_2\pi}$
on $s$ is shown in Fig.~\ref{repivpij}.

\subsection{Axial vector-pseudoscalar loop}
~

The axial vector-pseudoscalar meson state AP$=a_1(1260)\pi$ is
considered to be one of the states contributing to the four-pion
production amplitude \cite{pdg}. For soft pions, when taking into
account the requirements of chiral symmetry, this amplitude and
the corresponding partial width, are very complicated
\cite{ach00a,ach00b,ach08a,ach08b,czyz08,czyz01,ecker89}. This
prevents one from using the dispersion relation to obtain the
contribution of the four-pion state to the polarization operator
of the state $\rho_i$. Hence, in the present work, the simplest
$a_1\pi$ dominance model of the four-pion \cite{cmd4pi} production
is used: $e^+e^-\to\rho_i\to a_1\pi\to4\pi$. The amplitude of the
transition $\rho_i\to a_1\pi$ is chosen in the simplest form
\begin{eqnarray}
A(\rho_{iq}\to
a_{1k}\pi_p)&=&g_{\rho_ia_1\pi}[(\epsilon_{a_1}\epsilon_{\rho_i})(kq)-\nonumber\\&&
(\epsilon_{a_1}q)(\epsilon_{\rho_i}k)],\label{rhoia1pi}\end{eqnarray}
where $q$, $k$, and $p$ are, respectively, the four-momenta of the
mesons $\rho_i$, $a_1$, and $\pi$, while $\epsilon_{a_1}$ and
$\epsilon_{\rho_i}$ denote the polarization four-vectors of $a_1$
and $\rho_i$. The expression (\ref{rhoia1pi}) is chosen on the
grounds that it is explicitly transverse in the $\rho_i$ leg.

\begin{figure}
  \includegraphics[width=8cm]{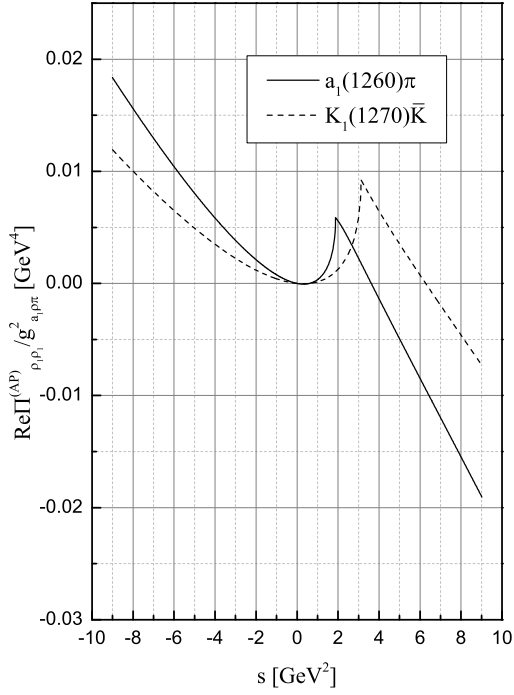}
  \caption{\label{repiap}The dependence of Re$\Pi^{(AP)}_{\rho_1\rho_1}(s)$ on energy squared
  $s$ for the  $AP=a_1(1260)\pi$ and $AP=K_1(1270)\bar K$ loops  demonstrating discontinuity of
  $\frac{d{\rm Re}\Pi^{(AP)}}{ds}$ at $s=(m_A+m_P)^2$. The
  parameters
  are $m_\rho=777$ MeV, $m_{a_1}=1230$ MeV, $m_{K_1}=1270$ MeV, $s_0=0.09$ GeV$^2$.}
\end{figure}

The diagonal polarization operators due to the AP loop are
represented in the form
\begin{equation}
\Pi_{\rho_i\rho_i}^{(AP)}=g^2_{\rho_iAP}\Pi^{(AP)},\end{equation}
where the quantity
$\Pi^{(AP)}\equiv\Pi^{(AP)}(s,m_{\rho_i},m_A,m_P)$ is calculated
from the dispersion relation
\begin{eqnarray}
\frac{\Pi^{(AP)}}{s}&=&\frac{1}{48\pi^2}
\int_{(m_A+m_P)^2}^\infty\frac{q_{AP}(s^\prime,m_A,m_P)}{(s^\prime)^{3/2}(s^\prime-s-i\varepsilon)}\times
\nonumber\\&&[(s^\prime+m^2_A-m^2_P)^2+2s^\prime
m^2_A]\times\nonumber\\&&\left(\frac{s_0+m^2_{\rho_i}}{s_0+s^\prime}\right)ds^\prime,
\label{dispAP}
\end{eqnarray}
where the expression
\begin{eqnarray}\label{widroiAP} \Gamma_{\rho_i\to
AP}(s)&=&\frac{g^2_{\rho_iAP}}{48\pi
s}\left[(s+m^2_A-m^2_P)^2+2sm^2_A\right]\times\nonumber\\&&q_{AP}(s,m_A,m_P)\left(\frac{s_0+m^2_{\rho_i}}{s_0+s}\right)
\end{eqnarray}
for the $\rho_i\to AP$ decay width found from the effective vertex
Eq.~(\ref{rhoia1pi}) is inserted into the integrand of the
dispersion relation. The result of integration is represented in
the form
\begin{eqnarray}\label{PiAP}
\Pi^{(AP)}&=&\frac{s}{96\pi^2}\left(\frac{m_-}{m_+}\right)^3\frac{m^2_{\rho_i}+s_0}{s+s_0}
\left[J(s)-\right.\nonumber\\&&\left.{\rm
Re}J(m^2_{\rho_i})\frac{s+s_0}{m^2_{\rho_i}+s_0}\right],
\label{piAP1}
\end{eqnarray}
where $m_\pm=m_A\pm m_P$,
\begin{eqnarray}
J(s)&=&sf(s)\left[\left(\frac{m^2_+}{s}-1\right)^2+\frac{2m_A}{m_-}\times\right.\nonumber\\&&
\left.\left(2+\frac{m_A}{m_-}\right)
\left(\frac{m^2_+}{s}-1\right)+6\frac{m^2_A}{m^2_-}\right]-\nonumber\\&&
s_0f_2\left[\left(\frac{m^2_+}{s_0}+1\right)^2-\frac{2m_A}{m_-}\times\right.\nonumber\\&&
\left.\left(2+\frac{m_A}{m_-}\right)
\left(\frac{m^2_+}{s_0}+1\right)+6\frac{m^2_A}{m^2_-}\right].
\label{Js}\end{eqnarray} The function $f(s)$ looks as
\begin{widetext}
\begin{eqnarray}
f(s)&=&\frac{m_+}{sm_-}\left\{(s-m^2_-)\frac{m_+}{m_-}\ln\frac{m_A}{m_P}-
s\left(\frac{m^2_+-m^2_-}{2m_+m_-}\ln\frac{m_A}{m_P}-1\right)+
\sqrt{|(s-m^2_+)(s-m^2_-)|}\times\right.\nonumber\\&&\left.
\left[\theta(m^2_--s)\ln\frac{\sqrt{m^2_+-s}+\sqrt{m^2_--s}}{\sqrt{m^2_+-s}+\sqrt{m^2_--s}}-
2\theta(s-m^2_-)\theta(m^2_+-s)\arctan\sqrt{\frac{s-m^2_-}{m^2_+-s}}-\right.\right.\nonumber\\&&\left.\left.
\theta(s-m^2_+)\left(-i\pi+\ln\frac{\sqrt{s-m^2_-}+\sqrt{s-m^2_+}}{\sqrt{s-m^2_-}-\sqrt{s-m^2_+}}\right)\right]\right\},
\label{fs}
\end{eqnarray}
\end{widetext}
and
\begin{widetext}
\begin{eqnarray}
f_2&=&-\frac{m_+}{m_-}\left[\frac{m_+}{m_-}\left(1+\frac{m^2_-}{s_0}\right)\left(\ln\frac{m_A}{m_P}-\frac{2m_-}{m_+}
\sqrt{\frac{m^2_++s_0}{m^2_-+s_0}}\tanh^{-1}\sqrt{\frac{m^2_-+s_0}{m^2_++s_0}}\right)
-\frac{m^2_+-m^2_-}{2m_+m_-}\ln\frac{m_A}{m_P}+1\right]\label{f2}
\end{eqnarray}
\end{widetext}
The dependence of Re$\Pi^{(AP)}_{\rho_1\rho_1}(s)$ on energy
squared is shown in Fig.~\ref{repiap}.

\begin{figure}
  \includegraphics[width=8cm]{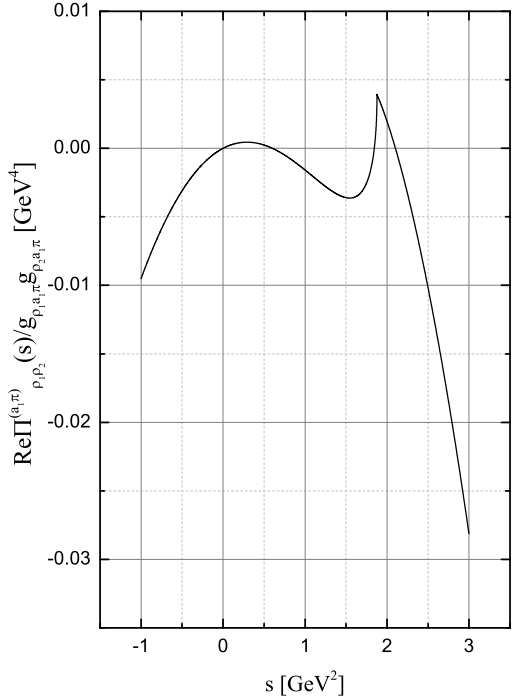}
  \caption{\label{repiap12}The dependence of Re$\Pi^{(a_1\pi)}_{\rho_1\rho_2}(s)/g_{a_1\rho_1\pi}g_{a_1\rho_2\pi}$ on energy squared in case of
  $m_{\rho_1}=770$ MeV and $m_{\rho_2}=1450$ MeV.}
\end{figure}

The function $G^{(AP)}(s)$ [Eq.~(\ref{Gab})]  necessary for
evaluation of the nondiagonal polarization operator due to the AP
loop is
\begin{equation}\label{GAP}
G^{(AP)}(s)=\frac{s^2}{48\pi^2}\left(\frac{4m_Am_P}{m_+m_-}\right)^3I^{(AP)}(s),\end{equation}
where
\begin{widetext}
\begin{eqnarray}\label{IAP}
I^{(AP)}(s)&=&\left\{\frac{1}{16}+\frac{m^2_-}{32m_Am_P}\left(\frac{m^2_+}{s}-1\right)+\frac{m_A+2m_-}{16m_P}
+\frac{1}{2}\left(\frac{m_+m_-}{4m_Am_P}\right)^2\left(\frac{m^2_+}{s}-1\right)\left(1-\frac{m^2_-}{s}\right)-\right.
\nonumber\\&&\left.\frac{m^2_-(m_A+2m_P)}{16m_Am^2_P}-\frac{3m^2_-}{16m^2_P}+\frac{m^2_+}{4m_Am_P}\left(1-\frac{m^2_-}{s}\right)
\left[\left(\frac{m^2_-}{4m_Am_P}\right)^2\left(\frac{m^2_+}{s}-1\right)^2+\right.\right.
\nonumber\\&&\left.\left.\frac{m^2_-}{2m_Am_P}\left(\frac{m^2_+}{s}-1\right)
+\frac{3m^2_-}{8m^2_P}\right]\right\}\times\frac{1}{2}\ln\frac{m_A}{m_P}+\frac{m_+m_-}{4m_Am_P}\left\{\frac{1}{6}
\left(\frac{m^2_+}{4m_Am_P}\right)^2+\frac{m^2_-}{4m_Am_P}\left[\frac{1}{24}-\right.\right.\nonumber\\&&\left.\left.
\frac{m^2_-}{16m_Am_P}\left(\frac{m^2_+}{s}-1\right)
-\frac{m_A+2m_-}{8m_P}\right]-\frac{1}{16}+\frac{3m^2_-}{32m_Am_P}\left(\frac{m^2_+}{s}-1\right)+\frac{3(m_A+2m_-)}{16m_P}
-\right.\nonumber\\&&\left.
\frac{1}{2}\left(\frac{m_+m_-}{4m_Am_P}\right)^2\left(\frac{m^2_+}{s}-1\right)\left(1-\frac{m^2_-}{s}\right)
-\frac{m^2_+(m_A+2m_P)}{16m_Am^2_P}\left(1-\frac{m^2_-}{s}\right)+\frac{3m^2_-}{16m^2_P}+\frac{1}{s}\times\right.\nonumber\\&&\left.
\left[\left(\frac{m^2_-}{4m_Am_P}\right)^2\left(\frac{m^2_+}{s}-1\right)^2+\frac{m^2_-}{2m_Am_P}\left(\frac{m^2_+}{s}-1\right)
+\frac{3m^2_-}{8m^2_P}\right]\left|(s-m^2_-)(s-m^2_+\right|^{1/2}\times\right.\nonumber\\&&\left.\left[
\frac{1}{2}\theta(m^2_--s)\ln\frac{\sqrt{m^2_+-s}+\sqrt{m^2_--s}}{\sqrt{m^2_+-s}-\sqrt{m^2_--s}}-\theta(s-m^2_-)\theta(m^2_+-s)
\arctan\sqrt{\frac{s-m^2_-}{m^2_+-s}}+\right.\right.\nonumber\\&&\left.\left.
\frac{1}{2}\theta(s-m^2_+)\left(i\pi-\ln\frac{\sqrt{s-m^2_-}+\sqrt{s-m^2_+}}
{\sqrt{s-m^2_-}-\sqrt{s-m^2_+}}\right)\right]\right\}.
\end{eqnarray}
\end{widetext}
The dependence of
Re$\Pi^{(a_1\pi)}_{\rho_1\rho_2}/g_{a_1\rho_1\pi}g_{a_1\rho_2\pi}$
on $s$ is shown in Fig.~\ref{repiap12}.

\subsection{Polarization operators used in fits}
~

Although the nature of the higher resonances $\rho(1450)$,
$\rho(1700)$, ... is the subject of current and future studies,
the quark-antiquark model relations between their coupling
constants are assumed:
\begin{eqnarray}\label{coupconst}
g_{\rho_iK^\ast K}&=&\frac{1}{2}g_{\rho_i\omega\pi},\nonumber\\
g_{\rho_iK_1(1270)K}&=&\frac{1}{2}g_{\rho_ia_1(1260)\pi}.
\end{eqnarray} Polarization operators which take into account
three channels described above are the following. The full
diagonal polarization operators are
\begin{equation}
\Pi_{\rho_i\rho_i}=\Pi^{(PP)}_{\rho_i\rho_i}+\Pi^{(VP)}_{\rho_i\rho_i}
+\Pi^{(AP)}_{\rho_i\rho_i}.\label{Pifulldia}\end{equation} In the
present work, we take into account the following analytically
calculated loops. First, we use the PP $\pi^+\pi^-$ and
$K^+K^-+K^0\bar K^0$ loops,
\begin{eqnarray}
\Pi^{(PP)}_{\rho_i\rho_i}&=&
g^2_{\rho_i\pi\pi}\left[\Pi^{(PP)}(s,m_{\rho_i},m_\pi)+\right.\nonumber\\&&\left.
\frac{1}{2}\Pi^{(PP)}(s,m_{\rho_i},m_K)\right],\end{eqnarray}with
$\Pi^{(PP)}$ given by Eq.~(\ref{PiPP1}). Second, we use  the VP
$\omega\pi^0$ and $K^\ast\bar K^+\bar K^\ast K$ loops,
\begin{eqnarray}
\Pi^{(VP)}_{\rho_i\rho_i}&=&
g^2_{\rho_i\omega\pi}\left[\Pi^{(VP)}(s,m_{\rho_i},m_\omega,m_\pi)+\right.\nonumber\\&&\left.
\Pi^{(VP)}(s,m_{\rho_i},m_{K^\ast},m_K)\right],\end{eqnarray}with
$\Pi^{(VP)}\equiv\Pi^{(VP)}(s,m_{\rho_i},m_V,m_P)$ given by
Eq.~(\ref{PiVP}).  Third, we use the AP
$a_1(1260)^+\pi^-+a_1(1260)^-\pi^+$, $K_1(1270)\bar K+$ c.c.
loops,
\begin{eqnarray}
\Pi^{(AP)}_{\rho_i\rho_i}&=&2g^2_{\rho_ia_1\pi}\left[\Pi^{(AP)}(s,m_{\rho_i},m_{a_1},m_\pi)+
\right.\nonumber\\&&\left.
\frac{1}{2}\Pi^{(AP)}(s,m_{\rho_i},m_{K_1(1270)},m_K)\right],\end{eqnarray}
with $\Pi^{(AP)}\equiv\Pi^{(AP)}(s,m_{\rho_i},m_A,m_P)$ given by
Eq.~(\ref{PiAP}).

Similar expressions are used for the nondiagonal polarization
operators,
\begin{equation}
\Pi_{\rho_i\rho_j}=\Pi^{(PP)}_{\rho_i\rho_j}+\Pi^{(VP)}_{\rho_i\rho_j}
+\Pi^{(AP)}_{\rho_i\rho_j},\label{Pifullij}\end{equation} where
\begin{widetext}
\begin{eqnarray}\label{Piabij}
\Pi^{(PP)}_{\rho_i\rho_j}&=&
g_{\rho_i\pi\pi}g_{\rho_j\pi\pi}\left[\Pi_{\rho_i\rho_j}(s,m_{\rho_i},m_{\rho_j},m_\pi,m_\pi)+
\frac{1}{2}\Pi_{\rho_i\rho_j}(s,m_{\rho_i},m_{\rho_j},m_K,m_K)\right],\nonumber\\
\Pi^{(VP)}_{\rho_i\rho_j}&=&
g_{\rho_i\omega\pi}g_{\rho_j\omega\pi}\left[\Pi_{\rho_i\rho_j}(s,m_{\rho_i},m_{\rho_j},m_\omega,m_\pi)+
\Pi_{\rho_i\rho_j}(s,m_{\rho_i},m_{\rho_j},m_{K^\ast},m_K)\right],\nonumber\\
\Pi^{(AP)}_{\rho_i\rho_j}&=&
2g_{\rho_ia_1\pi}g_{\rho_ja_1\pi}\left[\Pi_{\rho_i\rho_j}(s,m_{\rho_i},m_{\rho_j},m_{a_1},m_\pi)+
\frac{1}{2}\Pi_{\rho_i\rho_j}(s,m_{\rho_i},m_{\rho_j},m_{K_1},m_K)\right],
\end{eqnarray}
\end{widetext}
with $\Pi_{\rho_i\rho_j}(s,m_{\rho_i},m_{\rho_j},m_a,m_b)$ given
by Eq.~(\ref{Piij1}) in which the corresponding $G^{(ab)}(s)$ are
given in Eqs.~(\ref{GPP}), (\ref{GVP}), and (\ref{GAP}).

\section{Quantities for comparison with the data}\label{quantit}
~

The following reactions are considered in the present work:
\begin{equation}
e^+e^-\to\pi^+\pi^-,\label{pipi}
\end{equation}
\begin{equation}
e^+e^-\to\omega\pi^0,\label{ompi}
\end{equation}
and
\begin{equation}
e^+e^-\to\pi^+\pi^-\pi^+\pi^-.\label{4pi}
\end{equation}
The justification of the restriction to these reactions are given
in Introduction and  below in Sec.~\ref{results}. Let us turn to
the working expressions necessary for the comparison with the
experimental data.

\subsection{$\pi^+\pi^-$ production}
~

In the present case, the relevant quantity is the so-called bare
cross section of the reaction (\ref{pipi}):
\begin{equation}
\sigma_{\rm
bare}=\frac{8\pi\alpha^2}{3s^{5/2}}|F_\pi(s)|^2q^3_\pi(s)
\left[1+\frac{\alpha}{\pi}a(s)\right],\label{sigbare}
\end{equation}
where $F_\pi(s)$ is the pion form factor Eq.~(\ref{Fpi}), the
quantity $a(s)$ takes into account the radiation by the final
pions, and $\alpha=1/137$ is the fine-structure constant. The
necessary discussion concerning the quantities in
Eq.~(\ref{sigbare}) are given elsewhere \cite{ach11}.

\subsection{$\omega\pi^0$ production}
~
 The cross section of the reaction $e^+e^-\to\omega\pi^0$ is taken
 in the form

\begin{eqnarray}
\sigma_{e^+e^-\to\omega\pi^0}&=&\frac{4\pi\alpha^2}{3s^{3/2}}\left|A_{e^+e^-\to\omega\pi^0}
\right|^2\times\nonumber\\&&q^3_{\omega\pi}(s,m_\omega,m_\pi),\label{sigompi}\end{eqnarray}
where $q_{\omega\pi}$ is given by Eq.~(\ref{qab}),
\begin{eqnarray}
A_{e^+e^-\to\omega\pi^0}&=&(g_{\gamma\rho_1},g_{\gamma\rho_2},g_{\gamma\rho_3},\cdots)G^{-1}
\times\nonumber\\&&\left(%
\begin{array}{c}
  g_{\rho_1\omega\pi} \\
  g_{\rho_2\omega\pi} \\
  g_{\rho_3\omega\pi} \\
  \cdots\\
\end{array}%
\right),\label{aompi}\end{eqnarray}is the amplitude of the
reaction, and the matrix
\begin{equation}
G=\left(%
\begin{array}{cccc}
  D_1 & -\Pi_{12} & -\Pi_{13}&\cdots  \\
  -\Pi_{12} & D_2 & -\Pi_{23}&\cdots  \\
  -\Pi_{13} & -\Pi_{23} & D_3&\cdots  \\
  \cdots&\cdots&\cdots\\
  \end{array}%
\right)\label{G}\end{equation}is introduced in order to take into
account the strong mixing of the resonances $\rho_i$ \cite{ach11}.
Here, $\Pi_{ij}\equiv\Pi_{\rho_i\rho_j}$ are the polarization
operators. See Eqs.~(\ref{Pifulldia}) and (\ref{Pifullij}). The
nondiagonal $i\not=j$ terms describe the mixing. The inverse
propagators $D_i$ are given by Eq.~(\ref{repl}).

\subsection{$\pi^+\pi^-\pi^+\pi^-$ production}
~

The width of the decay $\rho_i\to2\pi^+2\pi^-$ in the model
(\ref{rhoia1pi}) is represented in the form
\begin{equation}
\Gamma_{\rho_i\to2\pi^+2\pi^-}(s)=g^2_{\rho_ia_1\pi}W_{\pi^+\pi^-\pi^+\pi^-}(s),\label{gam4pi}\end{equation}
where
\begin{eqnarray}
W_{\pi^+\pi^-\pi^+\pi^-}(s)&=&\frac{1}{12\pi}\int_{(3m_\pi)^2}^{(\sqrt{s}-m_\pi)^2}
\rho_{a_1}(m^2)\times\nonumber\\&&\left[\frac{(s+m^2-m^2_\pi)^2}{2s}+m^2\right]\times\nonumber\\&&
q_{a_1\pi}(s,m^2,m^2_\pi)dm^2\label{W4pi}\end{eqnarray}and
$q_{a_1\pi}$ is given by Eq.~(\ref{qab}). The function
\begin{equation}
\rho_{a_1}(m^2)=\frac{m_{a_1}\Gamma_{a_1}/\pi}{(m^2-m^2_{a_1})^2+m^2_{a_1}\Gamma^2_{a_1}}\label{rhoa1}\end{equation}
is introduced to take into account the large width of the
intermediate $a_1$ resonance in a minimal way, by taking the limit
of the fixed $a_1$ width. The mass  and  width of $a_1(1260)$ are
determined from fitting the data on the reaction
$e^+e^-\to\pi^+\pi^-\pi^+\pi^-$.

The cross section of the reaction $e^+e^-\to\pi^+\pi^-\pi^+\pi^-$
is represented in the form
\begin{widetext}
\begin{eqnarray}
\sigma_{\pi^+\pi^-\pi^+\pi^-}&=&\frac{(4\pi\alpha)^2}{3s^{3/2}}
\left|(g_{\gamma\rho_1},g_{\gamma\rho_2},g_{\gamma\rho_3},\cdots)G^{-1}\left(%
\begin{array}{c}
  g_{\rho_1a_1\pi} \\
  g_{\rho_2a_1\pi} \\
  g_{\rho_3a_1\pi} \\
  \cdots\\
\end{array}%
\right)\right|^2W_{\pi^+\pi^-\pi^+\pi^-}(s)
\label{sig4pi}\end{eqnarray} \end{widetext} with
$W_{\pi^+\pi^-\pi^+\pi^-}(s)$ given by Eq.~(\ref{W4pi}),
represents the effective phase-space volume of the four pions via
the  smearing [Eq.~(\ref{rhoa1})] of the $a_1\pi$ phase-space
volume. The matrix $G$ is the matrix of inverse propagators,
Eq.~(\ref{G}).

\section{Results of data fitting}\label{results}
~

This section is devoted to the presentation of the results of
fitting the data on the reactions $e^+e^-\to\pi^+\pi^-$,
$e^+e^-\to\omega\pi^0$, and $e^+e^-\to\pi^+\pi^-\pi^+\pi^-$. Two
possible fitting schemes were used.
\begin{itemize}
\item Scheme 1: Three resonances
$\rho(770)+\rho(1450)+\rho(1700)$ and the  PP loops of
pseudoscalar mesons are taken into account \cite{ach11,ach12}.
\end{itemize}

When adding the VP- and AP-loop contributions to the pion form
factor one should also include the VP and AP final states. These
states manifest, respectively, in the reactions
$e^+e^-\to\omega\pi^0$ and $e^+e^-\to\pi^+\pi^-\pi^+\pi^-$ which
should also be treated in the present framework. Since  energies
higher than 2 GeV are considered, the third heavy isovector
resonance $\rho(2100)$ is added. Hence, the scheme with the
resonances $\rho(770)+\rho(1450)+\rho(1700)+\rho(2100)$ and the
PP, VP, and AP loops in the polarization operators  is used. This
is scheme 2:
\begin{itemize}
\item Scheme 2. The data on the reactions $e^+e^-\to\pi^+\pi^-$ \cite{babar},
$e^+e^-\to\omega\pi^0$ \cite{snd13}, and
$e^+e^-\to\pi^+\pi^-\pi^+\pi^-$ \cite{bab4pi} are fitted
separately in the model which takes into account the resonances
$\rho(770)+\rho(1450)+\rho(1700)+\rho(2100)$ side by side while
allowing for the  PP, VP, and AP loops in the polarization
operators.
\end{itemize}
The parameters found from the fitting scheme 2 are listed in
Table~\ref{table1}.
%%%%%%%%%%%%%%%%%%%%%%%%%%%%%%%%%%%%%%%%%%%%%%%%%%%%%%%%%%%%%%%%%%%%%%%%%
\begin{table*}
\caption{\label{table1}The resonance parameters found from fitting
the data on the reactions $e^+e^-\to\pi^+\pi^-$ \cite{babar},
$e^+e^-\to\pi^+\pi^-\pi^+\pi^-$ \cite{bab4pi}, and
$e^+e^-\to\omega\pi^0$ \cite{snd13}, in the fitting scheme 2 (see
text). The parameter Re$\Pi^\prime_{\omega\rho}$ is responsible
for $\omega\rho$ mixing, see Ref.~\cite{ach11} for more detail.
The parameter $g_{\rho_4}$ can be found from the sum rule
$\frac{g_{\rho_1\pi\pi}}{g_{\rho_1}}+
\frac{g_{\rho_2\pi\pi}}{g_{\rho_2}}+\frac{g_{\rho_3\pi\pi}}{g_{\rho_3}}+
\frac{g_{\rho_4\pi\pi}}{g_{\rho_4}}=1$, which provides the correct
normalization $F_\pi(0)=1$.}
\begin{ruledtabular}
\begin{tabular}{llll}
 parameter&$e^+e^-\to\pi^+\pi^-$&$e^+e^-\to\pi^+\pi^-\pi^+\pi^-$&$e^+e^-\to\omega\pi^0$\\ \hline
 $m_{\rho_1}$[MeV]&$765.6\pm0.1$&$\equiv777$&$\equiv777$\\
 $g_{\rho_1\pi\pi}$&$6.336\pm0.004$&$5.78\pm0.01$&$6.63\pm0.03$\\
 $g_{\rho_1}$&$4.662\pm0.002$&$5.66\pm0.02$&$4.80\pm0.02$\\
 $g_{\rho_1a_1\pi}$[GeV$^{-1}$]&$2.37\pm0.05$&$0.26\pm0.02$&$5.61\pm0.08$\\
 $m_\omega$[MeV]&$782.02\pm0.10$&$\equiv782.02$&$\equiv782.02$\\
 Re$\Pi^\prime_{\omega\rho}$[GeV$^2$]&$(4.38\pm0.07)\times10^{-3}$&$-$&$-$\\  \hline
 $m_{\rho_2}$[MeV]&$1507\pm3$&$1122\pm1$&$1412\pm6$\\
 $g_{\rho_2\pi\pi}$&$-6.01\pm0.03$&$-5.45\pm0.03$&$-5.72\pm0.08$\\
 $g_{\rho_2}$&$78\pm2$&$8.74\pm0.04$&$21.6\pm.4$\\
 $g_{\rho_2\omega\pi}$[GeV$^{-1}$]&$6.66\pm0.07$&$38.2\pm0.2$&$32\pm2$\\
 $g_{\rho_2a_1\pi}$[GeV$^{-1}$]&$1.48\pm0.11$&$5.61\pm0.03$&$-1.37\pm0.15$\\
 \hline
 $m_{\rho_3}$[MeV]&$1831\pm4$&$1648\pm1$&$1661\pm5$\\
 $g_{\rho_3\pi\pi}$&$-2.800\pm0.008$&$3.681\pm0.005$&$-0.90\pm0.03$\\
 $g_{\rho_3}$&$8.45\pm0.05$&$4.302\pm0.005$&$4.90\pm0.04$\\
 $g_{\rho_3\omega\pi}$[GeV$^{-1}$]&$5.32\pm0.02$&$0.98\pm0.02$&$22.0\pm0.3$\\
 $g_{\rho_3a_1\pi}$[GeV$^{-1}$]&$0.41\pm0.06$&$-1.692\pm0.005$&$3.1\pm0.1$\\
 \hline
 $m_{\rho_4}$[MeV]&$2154\pm15$&$2390\pm5$&$1969\pm7$\\
 $g_{\rho_4\pi\pi}$&$-1.78\pm0.03$&$-4.5\pm0.1$&$-3.11\pm0.08$\\
 $g_{\rho_4\omega\pi}$[GeV$^{-1}$]&$8.65\pm0.12$&$11.39\pm0.05$&$8.19\pm0.10$\\
 $g_{\rho_4a_1\pi}$[GeV$^{-1}$]&$1.63\pm0.05$&$1.004\pm0.035$&$-0.91\pm0.02$\\
\hline
 $m_{K_1(1270)}$[MeV]&$\equiv1270$&$1265\pm1$&$1230\pm7$\\
 $m_{a_1(1260)}$[MeV]&$\equiv1230$&$1132\pm1$&$1268\pm4$\\
 $\Gamma_{a_1(1260)}$[MeV]&$-$&$1028\pm3$&$-$\\
  $s_0$[GeV]$^2$&$4.33\pm0.03$&$0.00037\pm0.00001$&$0.09\pm0.03$\\
  \hline
 $\chi^2/N_{\rm d.o.f}$&335/316&400/59&43/20\\
 \end{tabular}
\end{ruledtabular}
\end{table*}
%%%%%%%%%%%%%%%%%%%%%%%%%%%%%%%%%%%%%%%%%%%%%%%%%%%%%%%%%%%%%%%%%%%%%%%%%%
Let us comment on each of the  mentioned channels.

\subsection{Fitting $e^+e^-\to\pi^+\pi^-$ data}

\begin{figure}
\includegraphics[width=8cm]{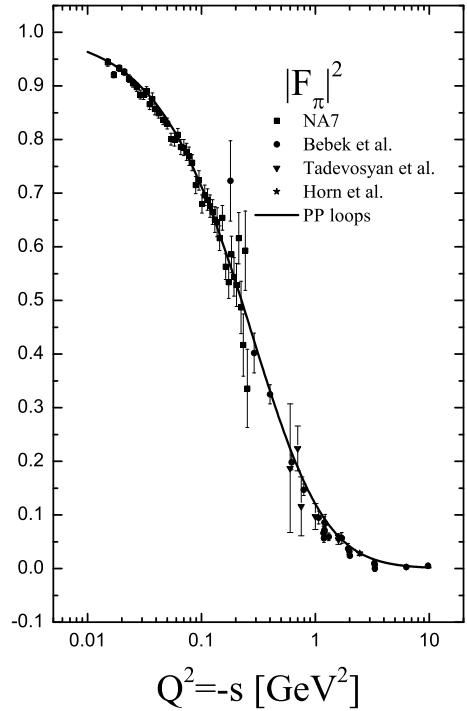}
\caption{\label{euclid}The pion form factor squared in the
space-like region $s<0$ evaluated using the resonance parameters
of the Table I in Ref.~\cite{ach11}, the BaBaR column, the fitting
scheme 1. See text. The experimental data are: NA7
\cite{amendolia}, Bebek {\it et al.} \cite{bebek}, Horn {\it et
al.} \cite{horn}, Tadevosyan {\it et al.}\cite{tadev}.}
\end{figure}

When fitting the data on the reaction $e^+e^-\to\pi^+\pi^-$ at
energies $\sqrt{s}\leq 1$ GeV in our previous publication
\cite{ach11,ach12,ach13}, the fitting scheme 1 was used. There,
the restriction to the PP loop was justifiable because of rather
low energies under consideration. Using the resonance parameters
found from fitting the data in the time-like region, the pion form
factor $F_\pi(s)$ in the space-like region $s<0$ was calculated up
to $-s=Q^2=0.2$ GeV$^2$ and compared with the NA7 data
\cite{amendolia}. A comparison  with the data
\cite{bebek,horn,tadev} in the wider range up to $-s=Q^2=10$
GeV$^2$ was made in Ref.~\cite{ach13}. In the present work, we
give the corresponding plot in Fig.~\ref{euclid} for the sake of
completeness.   The continuation to the space-like domain in the
fitting scheme 2 is discussed below.
%%%%%%%%%%%%%%%%%%%%%%%%%%%%%%%%%%%%%%%%%%%%%%%%%%%%%%%%%%%%%%%%%%%
%%%%%%%%%%%%%%%%%%%%%%%%%%%%%%%%%%%%%%%%%%%%%%%%%%%%%%%%%%%%%%%%%%%%

\begin{figure}
\includegraphics[width=10cm]{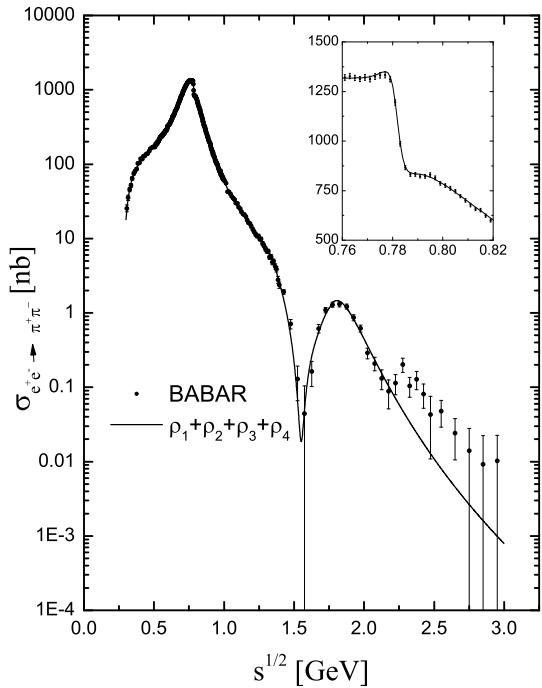}
\caption{\label{bab2pi}The cross section of the reaction
$e^+e^-\to\pi^+\pi^-$. The data are from Ref.~\cite{babar}, and
the curve is drawn using the resonance parameters of  scheme 2.
The $\rho-\omega$ resonance region is shown in the insert.}
\end{figure}

The cross section of the reaction $e^+e^-\to\pi^+\pi^-$  fitted in
the scheme 2 is shown in Fig.~\ref{bab2pi}. As for the $\rho(770)$
resonance parameters are concerned, one can observe that in
comparison with the fit in the scheme 1 \cite{ach11,ach12,ach13},
the bare mass of the resonance $\rho_1$ determined in the scheme 2
in the sensitive channel $e^+e^-\to\pi^+\pi^-$ is typically lower.
Compare Table \ref{table1} here and the Table I in, e.g.,
Ref.~\cite{ach11}. The same concerns the coupling constant
$g_{\rho_1}$ which parametrizes the leptonic decay width
(\ref{gamee}). The coupling constant $g_{\rho_1\pi\pi}$ in the
scheme 2 is greater than in the scheme 1. The above distinction
can be qualitatively explained by the effect of renormalization of
the coupling constants described in Ref.~\cite{ach11}. Indeed, as
was shown in Ref.~\cite{ach11}, the renormalization results in the
substitutions
\begin{eqnarray}
g_{\rho_1\pi\pi}&\to&Z_\rho^{-1/2}g_{\rho_1\pi\pi},\nonumber\\
g_{\rho_1}&\to&Z_\rho^{1/2}g_{\rho_1},\label{renorm}\end{eqnarray}
where
\begin{equation}
Z_\rho=1+\frac{d{\rm
Re}\Pi_{\rho_1\rho_1}(s)}{ds}\left|_{s=m^2_{\rho_1}}\right..
\label{Zrho}\end{equation}Equation (\ref{renorm}) means that the
bare $g_{\rho_1\pi\pi}$ obtained from the fit is related to the
"physical" one obtained from the visible peak, upon multiplying by
$Z^{1/2}_\rho$, while the opposite is true for $g_{\rho_1}$. The
contributions of the VP  loop to $d{\rm Re}\Pi_{\rho_1\rho_1}/ds$
near $s=m^2_{\rho_1}$, as is observed from Fig.~\ref{repivp}, is
positive and exceed the negative contribution from the PP loop,
see Fig.~7 in Ref.~\cite{ach11}. The same is true for the AP loop.
As a result, one has $Z_\rho>1$.

Although the energy behavior of the cross section up to
$\sqrt{s}=1.7$ GeV is described in the adopted model, including
the dip near 1.5 GeV, one can see that the structure in the
interval 2-2.5 GeV demands, in all appearance, additional
$\rho$-like resonances and/or  intermediate states in the loops.
We tried to include  the contribution of $K_1(1400)\bar K$+c.c.
states coupled solely to the resonance $\rho_4$, with the fitted
coupling constant $g_{\rho_4K_1(1400)K}$ and mass $m_{K_1(1400)}$.
This slightly improves the agreement in the interval
$1.75<\sqrt{s}<2$ GeV but  occurs at the expense of adding two
additional free parameters and does not result in reproducing the
peak near $\sqrt{s}=2.3$ GeV.
%%%%%%%%%%%%%%%%%%%%%%%%%%%%%%%%%%%%%%%%%%%%%%%%%%%%%%%%%%%%%%%%%%
%%%%%%%%%%%%%%%%%%%%%%%%%%%%%%%%%%%%%%%%%%%%%%%%%%%%%%%%%%%%%%%%%%%%

\begin{figure}
\includegraphics[width=8cm]{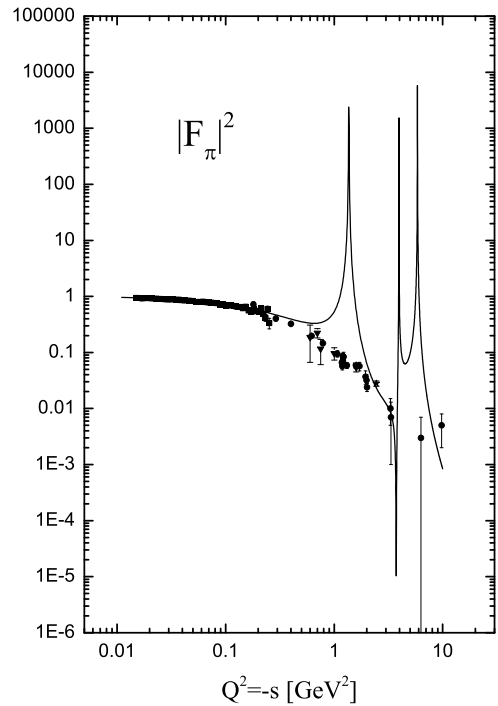}
\caption{\label{eucall}The same as in Fig.~\ref{euclid}, but with
the resonance parameters obtained in fitting scheme 2 (see text).}
\end{figure}

The continuation to the space-like region $s<0$ with the resonance
parameters obtained in the region $s>4m^2_\pi$ in  fitting scheme
2 with the VP and AP loops added, results in unwanted behavior of
$F_\pi(s)$, see Fig.~\ref{eucall}.  Specifically, the curve goes
through experimental points \cite{amendolia} up to $s=-0.2$
GeV$^2$, but at larger values of $-s=Q^2$ one encounters
infinities arising from the the Landau poles due to the VP and AP
loops.
%%%%%%%%%%%%%%%%%%%%%%%%%%%%%%%%%%%%%%%%%%%%%%%%%%%%%%%%%%%%%%%%%%
%%%%%%%%%%%%%%%%%%%%%%%%%%%%%%%%%%%%%%%%%%%%%%%%%%%%%%%%%%%%%%%%%%%%
As was pointed out in Ref.~\cite{ach11}, the Landau pole is
present even in the case of the PP loop, but its position is at
$\sqrt{Q^2}\approx 90$ GeV, that is, it is far from accessible
momentum transfers. In the case of the VP and AP loops the Landau
poles appear in the region accessible to existing experiments
\cite{bebek,horn,tadev}, because of the  large magnitude of the
coupling constant $g_{\rho_1\omega\pi}=13.2$ GeV$^{-1}$
\cite{fn1}.

An important feature of the new expression for the pion form
factor obtained in Ref.~\cite{ach11} which was not mentioned there
is that it does not require \cite{ach13} the commonly accepted
Blatt-Weisskopf centrifugal factor \cite{blatt}
$$C_\pi(k)=\frac{1+R^2_\pi k^2_R}{1+R^2_\pi k^2},$$ in the expression
for $\Gamma_{\rho\pi\pi}(s)$ \cite{pdg}. Here $k$ is the pion
momentum at some arbitrary energy while  $k_R$ is its value at the
resonance energy.  The fact is that the usage of $R_\pi$-dependent
centrifugal barrier penetration factor in particle physics -- for
example, in the case of the $\rho(770)$ meson \cite{pdg}, results
in the overlooked problem. Indeed, the meaning of $R_\pi$ is that
this quantity is the characteristic of the potential (or the
$t$-channel exchange in field theory) resulting in the phase
$\delta_{\rm bg}$ of the potential scattering in addition to the
resonance phase \cite{blatt}. For example, in case of the $P$-wave
scattering in the potential
$$U(r)=G\delta(r-R_\pi)$$ where the resonance scattering is
possible, the background phase is
$$\delta_{\rm bg}=-R_\pi k+\arctan(R_\pi k).$$  At the usual value
of $R_\pi\sim1$ fm, $\delta_{\rm bg}$ is not small. However, in
the $\rho$-meson region, the background phase shift $\delta_{\rm
bg}$ is negligible and the phase shift $\delta^1_1$ is completely
determined by the resonance; see Fig.~8 in Ref.~\cite{ach11}.
Therefore, the descriptions of the hadronic resonance
distributions which invoke  the parameter $R_\pi$, have a dubious
character.

\subsection{Fitting $e^+e^-\to\pi^+\pi^-\pi^+\pi^-$ data}
~

\begin{figure}
\includegraphics[width=9cm]{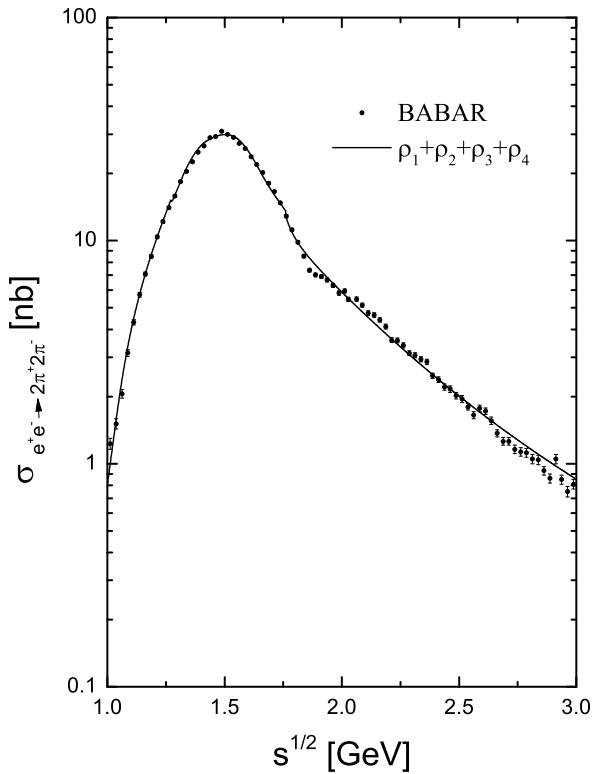}
\caption{\label{4pig}The cross section of the reaction
$e^+e^-\to\pi^+\pi^-\pi^+\pi^-$. The data are \cite{bab4pi}, the
curve is drawn using the resonance parameters of the scheme 2.}
\end{figure}

The dynamics of the reaction  $e^+e^-\to\pi^+\pi^-\pi^+\pi^-$ at
energies $\sqrt{s}<1$ GeV is determined by the chiral-invariant
mechanisms whose amplitudes are too cumbersome to include into the
loop integrations for the purposes of fits. Hence we restrict
ourselves by the energy range $1<\sqrt{s}<3$ GeV dominated by the
$\rho(1450)$, $\rho(1450)$, ... $s$-channel production mechanism.
The energy dependence of the $e^+e^-\to\pi^+\pi^-\pi^+\pi^-$
reaction cross section is shown in Fig.~\ref{4pig}. One can see
that at energies $\sqrt{s}>1.75$ GeV the chosen scheme with three
heavier rho-like resonances $\rho_{2,3,4}$ cannot reproduce the
structures in the measured cross section such as the bizarre sharp
turn in the energy behavior followed by fluctuations. As in the
case of the reaction $e^+e^-\to\pi^+\pi^-$, the contributions of
the AP loops $a_1(1260)\pi$ and $K_1(1270)\bar K$+c.c. coupled to
all $\rho_i$ resonances ($i=1,2,3,4$) were invoked to explain the
features above 1.75 GeV. The structures remain unexplained.
%%%%%%%%%%%%%%%%%%%%%%%%%%%%%%%%%%%%%%%%%%%%%%%%%%%%%%%%%%%%%%%%%%
%%%%%%%%%%%%%%%%%%%%%%%%%%%%%%%%%%%%%%%%%%%%%%%%%%%%%%%%%%%%%%%%%%%%

As is seen from  Table~\ref{table1}, the coupling constants
$g_{\rho_1}$ and $g_{\rho_1\pi\pi}$ of the $\rho(770)$ meson found
from fitting this channel, differ from those found from fitting
the $\pi^+\pi^-$ one. Furthermore, the coupling constant
$g_{\rho_1a_1\pi}$ found from fitting the channel
$e^+e^-\to\pi^+\pi^-\pi^+\pi^-$, is suppressed in comparison with
the naive chiral-symmetry estimate
$g_{a_1\to\rho_1\pi}\sim1/2f_\pi\sim5$ GeV$^{-1}$ \cite{ach05},
where $f_\pi=92.4$ MeV is the pion decay constant.  We believe
that this difference is an artifact of the oversimplified $a_1\pi$
model and the price that comes with  the possibility of using the
analytical calculation of the VP and AP loops to simulate the
contributions of the multiparticle meson states in polarization
operators \cite{fn3}. In the meantime, the coupling constants
$g_{\rho_1a_1\pi}\approx2.4$ and 5.6 GeV$^{-1}$ found from fitting
the $e^+e^-\to\pi^+\pi^-$ and $e^+e^-\to\omega\pi^0$ channels
respectively, look sensible. For comparison, the estimates of
$g_{\rho_1a_1\pi}$ in the model adopted in the present work are
$\sim6$ GeV$^{-1}$ and $\sim4$ GeV$^{-1}$, as extracted from
$\Gamma_{a_1}\approx0.6$ GeV and 0.3 GeV \cite{pdg}, respectively.
Note also that $\Gamma_{a_1\to\rho\pi\to3\pi}\sim1$ GeV when
evaluated  in the generalized hidden local symmetry chiral model
for $m_{a_1}\approx1.2$ GeV \cite{ach05}. The width of the visible
peak in Fig.~\ref{4pig} is about 0.44 GeV which should be compared
with $\Gamma_{\rho_3}(\sqrt{s}=m_{\rho_3})=0.45$ GeV evaluated
with the $\pi^+\pi^-\pi^+\pi^-$ column of the Table \ref{table1}.

\subsection{Fitting $e^+e^-\to\omega\pi^0$ data}
~

\begin{figure}
\includegraphics[width=9cm]{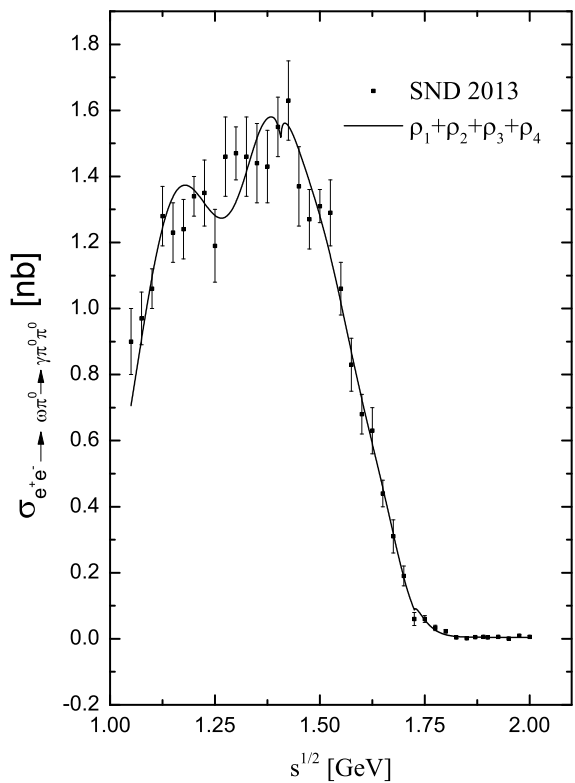}
\caption{\label{ompig}The cross section of the reaction
$e^+e^-\to\omega\pi^0\to\pi^0\pi^0\gamma$. The data are SND13
\cite{snd13}. The curve is drawn using the resonance parameters of
the scheme 2.}
\end{figure}

Quite recently, new  data  on the reaction $e^+e^-\to\omega\pi^0$
in the decay mode $\omega\to\pi^0\gamma$ were published by SND
collaboration \cite{snd13}. They are analyzed with the fitting
scheme 2. The resulting curve calculated with the parameters cited
in the Table~\ref{table1} is shown in Fig.~\ref{ompig}.
%%%%%%%%%%%%%%%%%%%%%%%%%%%%%%%%%%%%%%%%%%%%%%%%%%%%%%%%%%%%%%%%%%
%%%%%%%%%%%%%%%%%%%%%%%%%%%%%%%%%%%%%%%%%%%%%%%%%%%%%%%%%%%%%%%%%%%%

\section{Conclusion}\label{concl}
~

The main purpose of the present work is to describe the pion
electromagnetic form factor $F_\pi(s)$ up to the $J/\psi$ energy
range, using the expression obtained in Ref.~\cite{ach11}. This
expression, when restricted to the PP loops in polarization
operators, permits a good description of the data of SND, CMD-2,
KLOE, and $BABAR$ on $\pi^+\pi^-$ production in $e^+e^-$
annihilation at $\sqrt{s}<1$ GeV, describes the scattering
kinematical domain up to $-s=Q^2=10$ GeV$^2$, and does not
contradict the data on $\pi\pi$ scattering phase $\delta^1_1$. The
goal of extending the description to the energies up to 3 GeV in
the time-like domain was reached by the inclusion of the VP- and
AP-loops in addition to the PP ones. These loops contain the
couplings of the $\rho$-like resonances with the VP- and AP-states
and generate, in turn, the final states $\omega\pi^0$ and
$\pi^+\pi^-\pi^+\pi^-$ in $e^+e^-$ annihilation. Therefore,
consistency demands the treatment of these final states as well.
As is shown in the present work, the energy behavior of the cross
sections of the reactions $e^+e^-\to\omega\pi^0$ and
$e^+e^-\to\pi^+\pi^-\pi^+\pi^-$ obtained in the adopted simplified
model, does not contradict the data. The statistically poor
description of the cross section of the reaction
$e^+e^-\to\pi^+\pi^-\pi^+\pi^-$ is, probably, an artifact of the
oversimplified model for its amplitude which ignores both the
requirements of the chiral symmetry at lower energies and a
complicated intermediate state at higher energies. The proper
treatment of the reaction $e^+e^-\to\pi^+\pi^-\pi^+\pi^-$ is
beyond the scope of the present work. Nevertheless, we included
this poor description for the consistency of the presentation.

One should not wonder at the fact that the masses of heavier
$\rho$-like resonances quoted in Table \ref{table1} differ from
the values quoted in Ref.~\cite{pdg}. In fact, the values in
Ref.~\cite{pdg} are only educated guesses, and the masses of
heavier $\rho$-like resonances quoted by the Particle Data Group
fall into  wide intervals; for instance, $m_{\rho_2}=1265-1580$
MeV, and $m_{\rho_3}=1430-1850$ MeV \cite{pdg}. Furthermore, the
quoted values are usually obtained from fitting the data with the
help of the simplest parametrization such as the sum of the
Breit-Wigner amplitudes. In the meantime it is known that the
residues of the simple pole contributions not necessarily reveal
the true nature of the resonances involved in the process
\cite{ach07a,ach11a,ach12a} when the mixings and the dynamical
effects like the final-state interaction become essential.

The real problem is that the continuation to the space-like domain
of the expression for $F_\pi(s)$ with the contributions of the VP
and AP loops meets the difficulty of encountering the Landau
poles. By all appearances, this is the consequence of the chosen
parametrization of the vertex form factor which restricts the
growth of the partial widths as the energy increase, in a modest
way. A  stronger suppression could effectively suppress the
couplings of rho-like resonances   with the VP and AP states and,
in turn, push the Landau zeros to higher space-like momentum
transfers. This is the topic of a separate study.

We are grateful to M.~N.~Achasov for numerous discussions which
stimulated the present work. This  work is supported in part by
the Russian Foundation for Basic Research Grant no. 13-02-00039
and the Interdisciplinary project No 102 of the Siberian Division
of the Russian Academy of Sciences.

%%%%%%%%%%%%%%%%%%%%%%%%%%%%%%%%%%%%%%%%%%%%%%%%%%%%%%%%%%


\begin{thebibliography}{99}
\bibitem{fn2}
We use the opportunity to correct the misprint in the expression
for $\Pi^{(PP)}_1$ in Refs.~\cite{ach12,ach13} where the braces
were omitted when typesetting.
\bibitem{fn1}
The coupling constant $g_{\rho_1\omega\pi}$ is extracted from the
partial width of the decay $\omega(782)\to\pi^+\pi^-\pi^0$.
\bibitem{fn3}
Of course, more sophisticated  models which take into account
other possible mechanisms of the four-pion production in $e^+e^-$
annihilation and tau lepton decay (see for example
\cite{czyz08,czyz01,ecker89}) result in a better agreement with
the data. Nvertheless, we give here the results of fitting the
$e^+e^-\to\pi^+\pi^-\pi^+\pi^-$ data in the scheme with the
simplest quasi-two-body mechanism because it is this mechanism
which is used here in modelling the pion form factor. We believe
that it is appropriate to investigate the consequences of the
model used for the form factor (which is the main goal of the
present study) for other processes. A proper description of the
four-pion production cross section  is a separate problem.
\end{thebibliography}
\end{document}